\documentclass[conference,letterpaper]{IEEEtran}
\usepackage[utf8]{inputenc} 
\usepackage[T1]{fontenc}
\usepackage{url}
\usepackage{ifthen}
\usepackage{cite}
\usepackage[cmex10]{amsmath} % Use the [cmex10] option to ensure complicance
\usepackage{psfrag}

\usepackage{hyperref}       % hyperlinks
\usepackage{booktabs}       % professional-quality tables

\usepackage{bm}             % bold math
\usepackage{bbm}

\usepackage{amsfonts}
\usepackage{amscd}

\usepackage{mathtools}
\usepackage{mathrsfs}

\usepackage{tikz}
\usepackage{xcolor}

\usepackage{nicefrac}       % compact symbols for 1/2, etc.
\usepackage{microtype}      % microtypography

\usepackage{cleveref}

\usepackage{times}
\usepackage{xspace}
\usepackage{enumitem}

\usepackage{stmaryrd}
\usepackage{comment}
\newcommand{\family}{\mathcal{S}}

\newcommand{\RDs}{R^{\mathrm{w}}(D)} % Rate Distortion Function

\newcommand{\RDg}[1]{R_{P_\circ}(D, #1)} % Rate 
\newcommand{\Cs}[1]{C_{P_\circ}(B, #1)} % Rate Distortion Function
\newcommand{\EP}{P_{\hat{X}|X}} % Encoding Function

\newcommand{\BW}{\operatorname{BW}} %Bures-wasserstein distance
\newcommand{\Was}{\operatorname{W_2}}

\newtheorem{theorem}{Theorem}[section]

\newtheorem{lemma}[theorem]{Lemma}

\newcommand{\defeq}{\coloneqq}                      %LHS defined as RHS
\newcommand{\+}{\!+\!}
\renewcommand{\-}{\!-\!}
\renewcommand{\=}{\!=\!}

\renewcommand{\>}{\!>\!}

\newcommand{\ie}{\textit{i.e.}}

\newcommand{\R}{\mathbb{R}}     %real numbers
\newcommand{\Sym}{\mathbb{S}}   %set of symmetric real matrices
\newcommand{\Prob}{\mathcal{P}_{2+\epsilon}}
\newcommand{\Probc}{\mathcal{P}_{2}}

\newcommand{\cl}[1]{\left\{{#1}\right\}}        %curly bracket

\newcommand{\abs}[1]{\vert{#1}\vert}                    %absolute value
\newcommand{\norm}[2][\text{}]{\Vert{#2}\Vert_{#1}}     %norm of #2 in [#1]
\newcommand{\clf}[1]{\mathcal{#1}} %calligraphic

\newcommand{\tr}{\operatorname{tr}}         %trace
\newcommand{\E}{\operatorname{\mathbb{E}}} %expectation operator
\newcommand{\Cov}{\operatorname{Cov}} %covariance

\newcommand{\sampled}[1][\text{}]{\stackrel{#1}{\sim}}

\newcommand{\normal}{\operatorname{\mathcal{N}}} %normal distribution
\newcommand{\beq}[1]{\begin{align*}\label{eq:#1}}
\newcommand{\eeq}{\end{align*}}

\newcommand{\e}{\mathrm{e}}             %e number
\newcommand{\thalf}{\tfrac{1}{2}} %one half
\newcommand{\half}{\frac{1}{2}} %one half

\makeatletter
\newcommand{\xMapsto}[2][]{\ext@arrow 0599{\Mapstofill@}{#1}{#2}}
\def\Mapstofill@{\arrowfill@{\Mapstochar\Relbar}\Relbar\Rightarrow}
\makeatother

\begin{document}
%\title{Distributionally Robust Lossy Compression using the Wasserstein-2 Metric} 
% \title{A Distributionally Robust Approach to Lossy Compression using the Wasserstein Distance}
\title{A Distributionally Robust Approach to Shannon Limits using the Wasserstein Distance}

%%% Single author, or several authors with same affiliation:
\author{%
 \IEEEauthorblockN{Vikrant Malik, Taylan Kargin, Victoria Kostina, Babak Hassibi}
 \IEEEauthorblockA{\\Department of Electrical Engineering\\
                   Caltech\\
                   Pasadena, CA, USA\\
                   Email: \{vmalik, tkargin, vkostina, hassibi\}@caltech.edu}
}

%%% Several authors with up to three affiliations:
% \author{%
%   \IEEEauthorblockN{Andrew R.~Barron}
%   \IEEEauthorblockA{Department of Statistics and Data Science\\
%                     Yale University\\
%                     New Haven, CT, USA\\
%                     Email: andrew.barron@yale.edu}
%   \and
%   \IEEEauthorblockN{Claude E.~Shannon and David Slepian}
%   \IEEEauthorblockA{Bell Telephone Laboratories, Inc.\\ 
%                     Murray Hill, NJ, USA\\
%                     Email: \{csh, dsl\}@bell-labs.com}
% }

%%% Many authors with many affiliations:
% \author{%
%   \IEEEauthorblockN{Andrew R.~Barron\IEEEauthorrefmark{1},
%                     Claude E.~Shannon\IEEEauthorrefmark{2},
%                     David Slepian\IEEEauthorrefmark{2},
%                     and Jacob Ziv\IEEEauthorrefmark{2}\IEEEauthorrefmark{3}}
%   \IEEEauthorblockA{\IEEEauthorrefmark{1}%
%                    Department of Statistics and Data Science, Yale University, New Haven, CT, USA,
%                     andrew.barron@yale.edu}
%   \IEEEauthorblockA{\IEEEauthorrefmark{2}%
%                     Bell Telephone Laboratories, Inc.,
%                     Murray Hill, NJ, USA,
%                     \{csh,dsl,jz\}@bell-labs.com}
%   \IEEEauthorblockA{\IEEEauthorrefmark{3}%
%                     Department of Electrical Engineering, Technion---Institute of Technology, Haifa, Israel,
%                     jz@ee.technion.ac.il}
% }

\maketitle

%%%%%%
%% Abstract: 
%% If your paper is eligible for the student paper award, please add
%% the comment "THIS PAPER IS ELIGIBLE FOR THE STUDENT PAPER
%% AWARD." as a first line in the abstract. 
%% For the final version of the accepted paper, please do not forget
%% to remove this comment!
%%

\begin{abstract}
We consider the rate-distortion function for lossy source compression, as well as the channel capacity for error correction, through the lens of distributional robustness. We assume that the distribution of the source or of the additive channel noise is unknown and lies within a Wasserstein-2 ambiguity set of a given radius centered around a specified nominal distribution, and we look for the worst-case asymptotically optimal coding rate over such an ambiguity set. Varying the radius of the ambiguity set allows us to interpolate between the worst-case and stochastic scenarios using probabilistic tools. Our problem setting fits into the paradigm of compound source / channel models introduced by Sakrison \cite{sakrison1969rate} and Blackwell \cite{blackwell1959capacity}, respectively. This paper shows that if the nominal distribution is Gaussian, then so is the worst-case source / noise distribution, and the compound rate-distortion / channel capacity functions admit convex formulations with Linear Matrix Inequality (LMI) constraints. These formulations yield simple closed-form expressions in the scalar case, offering insights into the behavior of Shannon limits with the changing radius of the Wasserstein-2 ambiguity set.

\emph{Keywords-} distributionally robust compression, universal compression, Wasserstein distance, lossy source coding, rate-distortion function, distributionally robust coding, compound channel. 
\end{abstract}

\section{Introduction}
\subsection{Literature context}

Rapid technological advancements in recent years have spurred a significant increase in data production and usage. Major technological forces driving this data era include 5G/6G, the Internet of Things, machine learning, and autonomous vehicles. Core to these technologies are compression and reconstruction algorithms, which provide essential means for data storage and transmission.
For example, video streaming services require efficient data compression, yet the source distribution can vary greatly. Similarly, autonomous vehicles depend on sensor data, which is prone to distributional shifts that can lead to critical errors if not adequately addressed. In machine learning, models are often trained on a dataset that has undergone compression and decompression. These applications underscore the need for compression and error correction schemes that are resilient to distributional changes.

Variable-length universal codes that adapt their encoding rate to the source / noise distribution they observe have been a topic of longstanding interest in the information theory community. Perhaps the most famous of these is the Lempel-Ziv family of lossless codes  \cite{lz77, welch1984technique, ziv1978compression}
that learn the source distribution on the fly, while asymptotically achieving the entropy rate of the source. Lempel-Ziv codes have also been extended to apply to lossy compression \cite{yang1996simple,yang1997fixed}. An approach to universal lossy compression that utilizes Markov chain Monte Carlo with the reconstruction alphabet changing over time is proposed in \cite{baron2010mcmc} (continuous sources),\cite{jalali2012block} (finite alphabet sources). Variable-rate universal codes for channel coding are presented in \cite{tchamkerten2002feedback, mackay2005fountain, draper2004efficient, shulman2003communication, blits2012universal}.

The idea that enables the aforementioned coding strategies \cite{lz77, welch1984technique, ziv1978compression, yang1996simple,yang1997fixed, baron2010mcmc, jalali2012block, tchamkerten2002feedback, mackay2005fountain, draper2004efficient, shulman2003communication, blits2012universal} is that of variable-length coding, i.e., the length of the codeword being transmitted is adapted to the data that is being seen. However, many physical systems and much of coding theory require all codewords to be of the same length. In that scenario, the relevant figure of merit is the worst-case performance over a family of source / noise distributions, known as \emph{compound} rate-distortion function / channel capacity-cost function, and the corresponding asymptotic fundamental limits are given by minimax convex optimization problems \cite{sakrison1969rate}, \cite{blackwell1959capacity}. In the context of compression, \cite{sakrison1970rate} considers the distributional class of stationary random processes with bounded $4^{th}$ order moment, whereas \cite{poor1982rate} examines a class of sources determined by spectral capacities. \textcolor{black}{In the context of error correction, \cite{blackwell1960capacities} extends the compound channel capacity formula of \cite{blackwell1959capacity} to random coding, to arbitrarily varying channels (AVCs), and to adversarial channels, \cite{root1968capacity} shows the worst-case capacity of a class of arbitrarily varying Gaussian channels, and \cite{gamal1979capacity} studies compound capacity of broadcast channels; see \cite{lapidoth1998reliable} for a survey.}

In this work, we employ Wasserstein-2 ($\Was$) distance as the measure of distributional shifts from a nominal distribution to define a distributional family. The $\Was$-distance between distributions $P_1, P_2$ on $\R^d$ is defined as \cite{villani_optimal_2009}
\begin{align}
    \Was(P_1,P_2)\defeq \left({\inf_{P_{XY} \in \Pi(P_1,P_2)} \mathbb{E} \left[ \norm{X 
    - Y}^2 \right]}  \right)^{\thalf},
\end{align}
where $\Pi(P_1,P_2)$ denotes the set of all joint distributions with marginals $P_1$ and $P_2$.

In contrast to other commonly used statistical distances such as total-variation distance, Kullback–Leibler 
% \vk{are you using these abbreviations later on? No need to define if not using}
divergence, Hellinger distance, the $\Was$-distance incorporates information from the geometric structure of the underlying domain, making it more suitable for handling structured real-world data. Furthermore, the $\Was$-distance accounts for the cost of transporting mass from one probability distribution to another. This makes the $\Was$-distance a particularly suitable tool for quantifying robustness in communication systems, as it reflects the potential impact of variations in signal distributions on the fidelity of communication. For instance, in scenarios where small perturbations in a signal could lead to disproportionately large errors, the $\Was$-distance provides a measure of how these perturbations affect system performance. This makes it a natural choice for modeling scenarios where the goal is to maintain high fidelity in the presence of uncertain or varying signal distributions, thereby enhancing the reliability and efficiency of communication systems under distributionally robust frameworks.

Owing to its geometric interpretability and tractable formulation, $\Was$-distance has recently gained popularity as a statistical distance in diverse fields such as control \cite{kargin2023wasserstein}, filtering \cite{shafieezadeh2018wasserstein}, and machine learning \cite{kuhn2019wasserstein}. In
\cite{al2023distributionally}, a controller that minimizes
% \vk{aims, but fails, or minimizes?}
the worst-case regret across all disturbance distributions within a $\Was$-ambiguity set is proposed. The controller's parameters are obtained via a solution to
% \vk{Is this what you mean?} 
a semi-definite program (SDP), which is formulated leveraging the tractability of $\Was$-distance. Unlike prior robust control approaches, such as $H_\infty$ \cite{hassibi1999indefinite}, where the controller is designed to minimize worst-case cost against an adversarially generated disturbance, and regret-optimal control \cite{goel2020regret}, where the controller is designed to minimize the regret with respect to a hypothetical optimal noncausal controller, the controller in \cite{al2023distributionally} is designed to 
% \vk{add citation(s)}
attain the desired level of robustness via adjusting the size of the ambiguity set. Additionally, \cite{al2023distributionally} demonstrates that the worst-case disturbance follows a Gaussian distribution if the nominal distribution is Gaussian. Subsequent contributions \cite{dr-ro-mf} and \cite{kargin2023wasserstein} extend the framework in \cite{al2023distributionally} to control scenarios involving measurement noise and infinite horizon, respectively. In the domain of lossy source coding, \cite{lei2021out} trains a deep neural network compressor that achieves distributional robustness by incorporating $\Was$ transportation cost into the optimization problem.
The $\Was$-distance has also been used to measure the perceptual quality of the decompressed message \cite{blau2019rethinking} using what is called the Rate-Distortion-Perception (RDP) function. The work \cite{serra2023computation} characterizes the analytical bounds for the Gaussian RDP function.

\subsection{Contributions}

We investigate the minimax rate-distortion function (RDF) formulated in \cite{sakrison1969rate} and the minimax capacity formulated in \cite{blackwell1959capacity} for an unknown source / noise distribution residing within a $\Was$-ambiguity set centered at a given nominal distribution. In the case of source compression, we assume that the source distributions in the ambiguity set have a finite $2\+\epsilon$-th moment for an arbitrarily small $\epsilon\>0$.
The finiteness of $(2\+\epsilon)$-th moment is a common assumption on the source distributions in rate-distortion theory to control the growth of the mean square error distortion (see \cite[Sec. 23.3]{polyanskiy2022information}).

In the case of a Gaussian nominal source distribution, we show that the worst-case source / noise distribution in the $\Was$ ambiguity set is also Gaussian. To do so, we leverage the Gelbrich bound \cite[Thm. 2.1]{gelbrich} on the $\Was$ distance and the Gaussian saddle point properties of mutual information \cite[Thm. 5.11]{polyanskiy2022information}. Our analysis reveals an expression of the minimax rate-distortion function / channel capacity solely in terms of the covariance matrices of distributions within the ambiguity set and the radius of the $\Was$-ball. Adjusting the radius of the $\Was$-ball enables a gradual interpolation from a known nominal source distribution (at a radius of 0) to an uncertain distribution as the radius increases.
In the scalar case, we derive closed-form expressions demonstrating the effect of the varying radius on those fundamental limits.

\textit{Notation.}
\textcolor{black}{The set of real numbers is denoted by $\R$. For a real-valued function $f$, we denote $f^{+}(\cdot) \defeq \max\{0, f (\cdot) \}$. For a vector $x \in \R^d$, we denote by $\norm{x}$, the Euclidean norm of $x$. For a matrix $A \in \R^{d \times m}$, its transpose is denoted by $A^T \in \R^{m \times d}$. For square matrices, $\tr(\cdot)$ and $\abs{\cdot}$ are the trace and the determinant operation. The notations $\Sym_{+}^d$ and $\Sym_{++}^d$ represent the sets of $d\times d$ symmetric, positive semidefinite and positive definite matrices, respectively. For two positive semidefinite matrices, $\succeq$ denotes the Löwner order. The set of all probability distributions over $\R^n$ with bounded $p^\textrm{th}$-moment is denoted by $\mathcal{P}_p(\R^n)$. For a random variable $X$ over $\R^n$, we denote its probability distribution by $P_X$. The conditional distribution of a random variable $Y$ given $X$ is denoted by $P_{Y|X}$. The expectation with respect to the distribution $P_X$ is denoted by $\E_{P_X}[\cdot]$. The Gaussian distribution with mean $\mu \in \R^d$ and covariance $\Sigma \in \Sym_{+}^d$ is denoted by  $\normal(\mu, \Sigma)$. The mutual information between random variables $X$ and $Y$ is denoted by $I(X;Y)$. The differential entropy of $X$ is denoted by $h(X)$, and the conditional differential entropy of $X$ given $Y$ is given by $h(X|Y)$.}

\section{Main results}
In this section, we state our main results$\colon$ Theorem \ref{th:main} shows an SDP for the worst-case RDF defined in \eqref{eq:wcrdf}, below, and Theorem \ref{th:main_cap} shows an SDP for the worst-case capacity, defined in \eqref{eq:wccap}.

% \end{align}

\subsection{Source Compression}
Fix $\epsilon > 0$. Consider a source that generates i.i.d. symbols $X_i \in \R^{d}$ following an unknown probability distribution $P_X \in \family$, where $\family \subseteq \Prob(\R^{d})$ is a distributional family. An $n$-block of source symbols $X^n = (X_1, \ldots, X_n)$ is mapped into one of the $\exp(nR)$ distinct codewords $\widehat{X}^n = (\widehat{X}_1, \ldots, \widehat{X}_n)$ while satisfying the average distortion constraint 
%$\frac 1 n \sum_i  \E\| X_i - \widehat X_i||^2 = d(X^{n}, \widehat{X}^{n}) \leq D$. 
$\frac 1 n \sum_{i=1}^n  \E[\norm{X_i-\widehat{X}_i}^2] \leq D$. 

The minimum universally achievable coding rate $R_{\family, n}(D)$ is defined as the minimum $R$ at which such a mapping exists, regardless of the $P_X \in \family$. The minimum asymptotically achievable universal coding rate is the operational compound RDF \cite{sakrison1969rate}, $R_{\family}(D) = \lim \sup_{n \to \infty} R_{\family, n}(D)$. Sakrison~\cite{sakrison1969rate} established the following single-letter formula for the compound RDF: 
\begin{equation}
    R_{\clf{S}}(D)= \inf_{\substack{\EP\colon\R^{d} \mapsto \R^d \\ \mathbb{E}\left[\|X - \widehat{X}\|^2\right]  \leq D}} \sup_{P_X \in \family} I(X; \widehat{X}).
    \label{eq:wcrdf}
\end{equation}
The achievability result of the coding theorem in \eqref{eq:wcrdf} assumes that the class of distributions $\family$ is compact, a notion defined by \cite{sakrison1969rate} as follows: a class of distributions $\family$ is compact if, for any $\epsilon > 0$, there is a totally bounded set $\mathcal{K}\subset \R^d$
and a function $f \colon \R^d \to \R^d$ with range $\mathcal{K}$ such that,
\begin{equation}
    \sup_{P_X\in \family}\mathbb{E}_{P_X}\left[\norm{X - f(X)}^2\right]\leq \epsilon.
\end{equation}
Any class of distributions with bounded $2 + \epsilon$ order moments for $\epsilon>0$ satisfies the compactness definition given in \cite{sakrison1969rate}. However, the distributions in the $\Was$-ball do not necessarily satisfy this condition. Therefore, to ensure that the function on the right side of \eqref{eq:wcrdf}, whose computation constitutes one of our main results, has an operational meaning, we consider only those distributions in the $\Was$-ball that have bounded $2 + \epsilon$ order moments.
% \eqref{eq:wass_ball_rdf}. 

\begin{comment}
    The worst case RDF defined in \eqref{eq:wcrdf} satisfies,
\begin{align}
\RDs = &\sup_{\Sigma_X \in \Wm} \nonumber\\ &\min _{\substack{A \in \mathbb{R}^{d \times d}, \Sigma_Z \succeq 0 \\ \tr((A - I)^T (A - I) \Sigma_X + \\ \Sigma_Z) \leq D}} \frac{1}{2} \log \left(\frac{\left|A \Sigma_X A^T+\Sigma_Z\right|}{\left|\Sigma_Z\right|}\right),
\end{align}
where $\Sigma_\circ$ is the covariance matrix of the nominal distribution $P_0$.
Moreover, worst case RDF given in \eqref{eq:wcrdf} is equal to the dual formulation given as,
\begin{align}
    \RDs &=  \sup_{P_X \in \Wr} \inf_{\substack{\EP \\ \mathbb{E}\left[\|X, \widehat{X} \|^2\right]  \leq D}} I(X; \widehat{X}).
\end{align}
where the $\sup$ is achieved by a Gaussian distribution.
% \label{th:main_rdf}
\end{comment}

Our main result for source compression computes the compound rate-distortion function \eqref{eq:wcrdf} for the $\Was$-ambiguity set of distributions
% $\family = \Wr$
$\family = \cl{P \in \Prob(\R^{d})  \mid  \Was(P,\, P_\circ) \leq r}$.
% \eqref{eq:wass_ball_rdf}
For brevity, we denote the compound rate-distortion function with such a choice of the ambiguity set by 
\begin{align}
   \RDg{r} \defeq R_{\family}(D). \label{eq:RDg} 
\end{align}
If $r = 0$, \eqref{eq:RDg} reduces to the rate-distortion function of the known distribution $P_\circ$. In our analysis, we assume a Gaussian nominal distribution.

To set the stage for our result, we denote the Bures-Wasserstein metric on the positive semi-definite cone $\Sym_+^d$ \cite{bhatia_bures-wasserstein_2017} as,
\begin{equation}
     \BW(\Sigma, \Sigma_\circ) \!\defeq \!\left(\tr(\Sigma) \+ \tr(\Sigma_\circ) \- 2\tr(\Sigma^{\half} \Sigma_\circ \Sigma^{\half} )\right)^{1/2}.
     \label{eq:bw_dist}
\end{equation} 
Moreover, note that the RDF of a $d-$dimensional multivariate Gaussian vector $X\sampled\normal(0,\Sigma)$ is given by \cite[Eqn. (13)]{kolmogorov1956shannon}$\colon$
\begin{equation}\label{eq:gaussian_rdf}
\begin{aligned}
& R_{\normal(0,\Sigma)}(D)= \inf _{ \substack{A \in \mathbb{R}^{d \times d}, \Sigma_Z \succeq 0 \colon \\
\tr( (A \- I) \Sigma(A \- I)^T \+ \Sigma_Z) \leq D
}
}
\frac{1}{2} \log  \frac{\left|A \Sigma A^T+\Sigma_Z\right|}{\left|\Sigma_Z\right|}.
\end{aligned}
\end{equation}
The reconstructed message $\widehat{X}$ follows the forward law $\widehat{X} = AX + Z$, where $Z\sampled\normal(0,\Sigma_Z)$ is independent of $X$. The solution to \eqref{eq:gaussian_rdf} is given by reverse-waterfilling on the eigenvalues \cite[Thm. 10.3.3]{tc}. 
If $d = 1$, the RDF in \eqref{eq:gaussian_rdf} takes the form \eqref{eq:scalar_rdf_tc}, below.

\begin{theorem}[Compound RDF for $\Was$ ambiguity set]
\label{th:main}
The compound RDF \eqref{eq:RDg} with a Gaussian center $P_\circ = \mathcal N \left( 0, \Sigma_\circ \right)$ 
is given as, 
\begin{equation}
    \begin{aligned}
\RDg{r} &\= \sup_{\substack{\Sigma \succeq 0, \\ \BW(\Sigma, \Sigma_\circ) \leq r}} R_{\normal(0,\Sigma_\circ)}(D),
\end{aligned}
\label{eq:main}
\end{equation}
where the function $\RDg{r}$ is achieved by a Gaussian $P_X$. 
\end{theorem}
\begin{IEEEproof}
 We first refer to \cite{sakrison1969rate} to establish strong duality, see Lemma~\ref{lem:rdfdual}, below. We then present an upper bound on the compound RDF and show that it is achieved by a Gaussian source distribution in the $\Was$-ball if the nominal source distribution $P_\circ$ is Gaussian. This extends the classical result \cite[Eqn. (13)]{kolmogorov1956shannon} showing that, among all distributions with a fixed second-order moment, a Gaussian source achieves the largest single source RDF (See Section~\ref{sec:mainproof} for details).
\end{IEEEproof}

\begin{lemma}[{Strong duality of the compound RDF}]
The compound RDF \eqref{eq:RDg} admits a dual formulation given by
\begin{equation}
\RDg{r}  \= \!\sup_{\substack{P_{X|X_\circ}\colon\R^{d} \mapsto \R^d \\ P_X \in \Prob(\R^{d}) \\ \mathbb{E}\left[\norm{X \- X_\circ}^2\right] \leq r^2 }} \inf_{\substack{\EP\colon\R^{d} \mapsto \R^d \\ \mathbb{E}\left[\norm{X - \widehat{X}}^2\right]  \leq D}} I(X; \widehat{X}),
\label{eq:rdfdual}
\end{equation}
where $X_\circ \sim P_\circ$.
\label{lem:rdfdual}
\end{lemma}
\begin{IEEEproof}
    The result follows via an application of \cite[Thm. Source Encoding]{sakrison1969rate} to mean square error and a $\Was$ ambiguity set.
\end{IEEEproof}
In \eqref{eq:rdfdual}, the nominal source $X_\circ \sim P_\circ$, the worst-case source $X$ and the reconstructed signal $\widehat{X}$ can be regarded as a Markov chain:
\begin{equation*}
    % \widehat{X} \xleftarrow{\EP\,} X \xleftarrow{P_{X\mid X_\circ}}  X_\circ.
    X_\circ \xrightarrow{P_{X|X_\circ}} X \xrightarrow{\EP\,} \widehat{X}.
\end{equation*}
We can interpret \eqref{eq:rdfdual} as a Nash equilibrium of a zero-sum game between two competing channels: while the channel $P_{X|X_\circ}$ adversarially maximizes the compression rate from $X$ to $\widehat{X}$ under the transportation cost $\mathbb{E}\norm{X \- X_\circ}^2 \leq r^2$ from the nominal source to "deceive" the encoder-decoder channel $\EP$, the channel $\EP$ minimizes the compression rate from $X$ to $\widehat{X}$ under the distortion constraint $\mathbb{E}\norm{X \- \widehat{X}}^2 \leq D$ by assuming the worst-case source from $P_{X|X_\circ}$.

To elucidate the tradeoff exposed in Theorem~\ref{th:main}, consider a scalar-source RDF with an unknown distribution from a $\Was$-ambiguity set centered at $P_\circ = \mathcal{N}(0, \sigma^2_\circ)$.
% \subsubsection{Scalar Case}
\label{sec:scalar_rdf}
First, recall the RDF of a scalar Gaussian source \cite[Eqn. (10.36)]{tc}:
\begin{align}
    R_{\normal(0,\sigma^2)}(D) = \frac 1 2 \log^{+} \frac{\sigma^2}{D}.
    \label{eq:scalar_rdf_tc}
\end{align}
Second, observe that the Bures-Wasserstein distance \eqref{eq:bw_dist} between scalars is,
\begin{align}
   \BW(\sigma^2, \sigma^2_\circ) = |\sigma - \sigma_\circ|.% \\
   \label{eq:bwsca}
\end{align}
We now apply Theorem \ref{th:main}. The compound RDF for the scalar case, using \eqref{eq:main}, is given as,
\begin{align}
    R_{\mathcal{N}(0, \sigma^2_\circ)} (D, r) &= \sup_{|\sigma - \sigma_\circ| \leq r } R_{\normal(0,\sigma^2)}(D)  \label{eq:rdfscalar_sup0} \\
     &= \sup_{|\sigma - \sigma_\circ| \leq r } \frac 1 2 \log^{+} \frac{\sigma^2}{D}  \label{eq:rdfscalar_sup} \\
    &= \frac 1 2 \log^{+} \frac{(\sigma_\circ + r)^2}{D}.
    \label{eq:rdfscalar}
\end{align}
The expression \eqref{eq:rdfscalar} is immediate due to the fact that the logarithm is monotonic.
Note that for $r = 0$, the compound RDF is the Shannon RDF for $P_\circ$, and for $r > 0$, it is the Shannon RDF for $P_X = \mathcal{N}(0, (\sigma_\circ + r)^2)$. Thus, as the radius of the ambiguity set increases, the required number of bits to achieve the same distortion increases.

\begin{figure}
    \centering
    \includegraphics[width=0.9\linewidth]{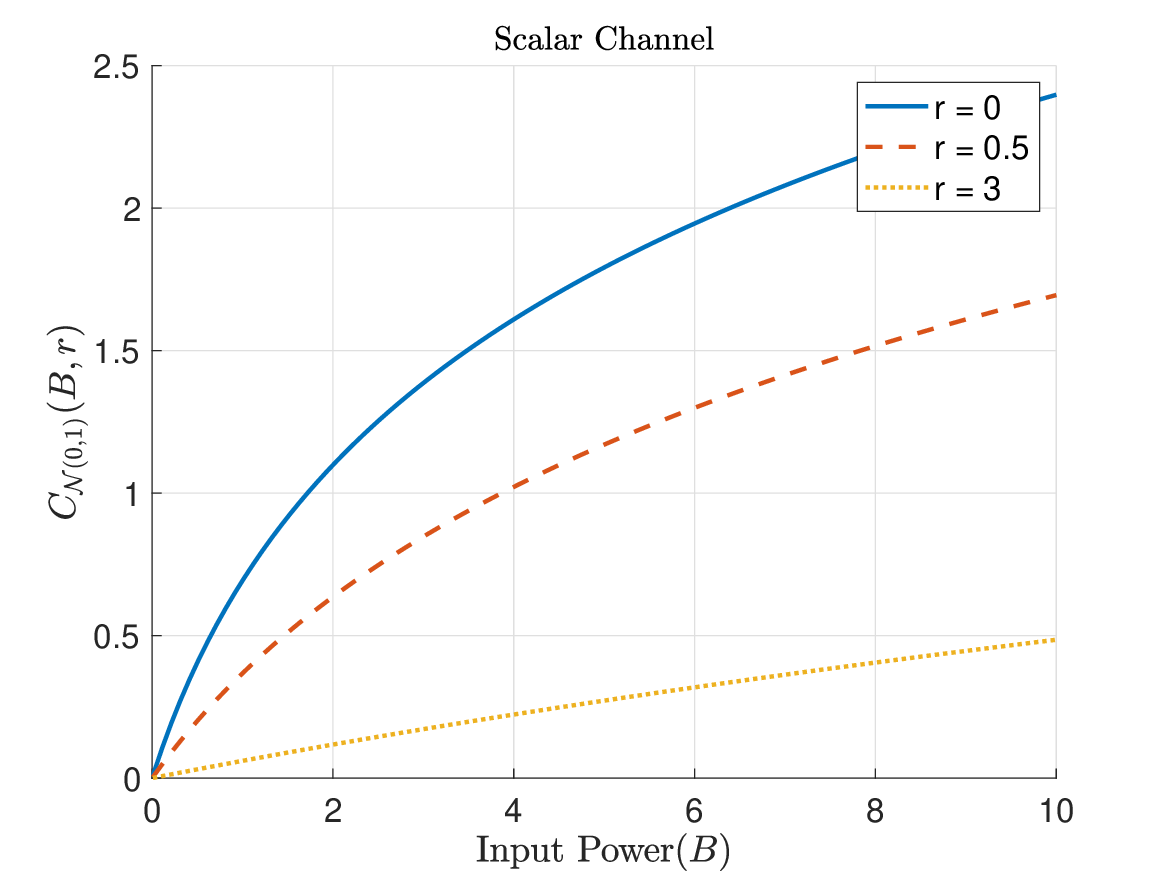}
    \caption{The scalar compound capacity for $P_{\circ} = \mathcal{N}(0,1)$ for $r = 0$. The scalar compound capacity for $r = 0$ is the Shannon capacity for $P_\circ$ and for $r > 0$ is the Shannon capacity for $P_Z = \mathcal{N}(0, (1 + r)^2)$. As the radius of the ambiguity set increases, the required input power $B$ to achieve the same capacity increases.}
    \label{fig:drccf}
\end{figure}

\subsection{Channel Coding}
\label{sec:channel_coding}
Consider transmission of an equiprobable message $W \in \left\{ 1, \ldots, \exp(nR) \right\}$ over an i.i.d. additive noise channel, which, upon receiving $X_i \in \R^{d}$, outputs $Y_i \in \R^{d}$, where $Y_i = HX_i + Z_i$, with $Z_i \sim P_Z$ independent of each other and of $X^i = (X_1, \ldots, X_i)$, and $H$ is fixed. Here, $P_Z \in \family$ is fixed but unknown. The encoder injectively maps $W$ to a codeword $X^n$ under the average power constraint $\frac 1 n \sum_{i = 1}^n  \mathbb E \| X_i \|^2\leq B$. The codeword is transmitted over the channel, and, upon receiving an $n$-block of channel outputs $Y^n \= (Y_1, \ldots, Y_n)$, the decoder outputs $\widehat{W}$, an estimate of $W$.

The maximum universally achievable coding rate $C_{\family, n}(B, \epsilon)$ compatible with average error probability $\mathbb{P}\left[W \neq \widehat{W}\right] \leq \epsilon$ is defined as the maximum rate $R$ for which such an encoder-decoder mapping exists, regardless of $P_Z \in \family$. The maximum asymptotically achievable universal coding rate is the operational compound capacity-cost function \cite{blackwell1959capacity}, $C_{\family, n}(B) = \lim_{\epsilon \to 0}\lim \inf_{n \to \infty} C_{\family, n}(B, \epsilon)$. Blackwell~\cite{blackwell1959capacity} established
a single-letter formula for the compound capacity-cost function of a discrete memoryless channel, which was extended to continuous alphabet domain in \cite{csiszar1992arbitrarily, loyka2016general}$\colon$
\begin{equation}
    C_{\clf{S}}(B)= \sup_{\substack{P_X \text{ on } \R^{d} : \\ \tr(\Sigma)  \leq B}} \inf_{\substack{P_Z \in \clf{S}}} I(X; HX + Z),
    \label{eq:wccap}
\end{equation}
where $\Sigma$ is the covariance of $X \sim P_X$.  Our main result for channel coding computes the compound capacity \eqref{eq:wccap} for the $\Was$-ambiguity set of additive channel noises 
\begin{align}
    \family = \cl{P \in \Probc(\R^{d})  \mid  \Was(P,\, P_\circ) \leq r}.
\end{align}
\textcolor{black}{Here, we assume that the nominal distribution $P_\circ \in \Probc(\R^{d})$ to ensure weak compactness of $\family$ \cite[Thm. 1]{yue2022linear}. This assumption entails $P \in \Probc(\R^{d})$ \cite[Lem. 1]{yue2022linear} for all $P$ in the $\Was$ ball.} For brevity, we denote the compound capacity-cost function with such a choice of the ambiguity set by 
\begin{align}
  \Cs{r}   \defeq C_{\family}(B) \label{eq:Cs}.
\end{align}
\textcolor{black}{Our problem setting satisfies the regularity conditions \cite[H1-H4]{csiszar1992arbitrarily}$\colon$
\begin{enumerate}
    \item The input and output alphabets as well as the set of states (which correspond to the probability measures in the $\Was$ ball equipped with the $\Was$ distance) are separable metric spaces, and the output alphabet is complete.
    \item The channel depends continuously on the input and states, \ie, for a sequence of inputs $X_n \rightarrow X$ and states $P_{Z_n} \rightarrow P_Z$, $P_{Y|X, P_Z}(\cdot | X_n, P_{Z_n})$ converges weakly to  $P_{Y|X, P_Z}(\cdot | X, P_Z)$. Here, $P_{Y|X, P_Z}(\cdot | X, P_Z)$ represents a channel with a given input $X$ and a fixed noise distribution $P_Z$ in the $\Was$ ball. This follows from the fact that the $\Was$-ball is weakly compact whenever the nominal distribution has a finite $2$-nd order moment \cite[Thm. 1]{yue2022linear}.
    \item The constraint function $\norm{\cdot}^2$ on the input is Borel measurable.
    \item There exists a sequence of input alphabets that satisfy the given power constraint.
\end{enumerate}}
The validity of the Blackwell formula \eqref{eq:wccap} ensures that the function on the right side of \eqref{eq:wccap}, whose computation constitutes one of our main results, has an operational meaning.

% Our main result for channel coding computes the compound capacity \eqref{eq:wccap} for the $\Was$-ambiguity set of additive channel noises $\family = \cl{P \in \Probc(\R^{d})  \mid  \Was(P,\, P_\circ) \leq r}$. \textcolor{red}{Here, we assume that the nominal distribution $P_\circ \in \Probc(\R^{d})$ to ensure weak compactness of the $\Was$ ambiguity set \cite[Thm. 1]{yue2022linear}. This assumption entails $P \in \Probc(\R^{d})$ \cite[Lem. 1]{yue2022linear}.} For brevity, we denote the compound capacity-cost function with such a choice of the ambiguity set by 
% \begin{align}
%   \Cs{r}   \defeq C_{\family}(B) \label{eq:Cs}.
% \end{align}
Note that the capacity-cost function for a $d$-dimensional multivariate Gaussian noise vector $Z_g \sim \mathcal{N}(0, \Sigma_Z)$ is given by (See \cite[Thm. 9.1]{el2011network} and Lemma \ref{lem:multi_cap} in Appendix \ref{sec:proof_cap} below), 
\begin{align}
    C_{\mathcal{N}(0, \Sigma_Z)}(B) = & \sup_{ \substack{\Sigma \succeq 0 \colon \\ \text{tr}(\Sigma) \leq B }} \log \frac{|\Sigma_Z + H\Sigma H^{T}|}{|\Sigma_Z|}.
\end{align}
\begin{theorem}[Compound capacity-cost function for $\Was$ ambiguity set]
\label{th:main_cap}

The compound capacity \eqref{eq:Cs} with a Gaussian center $P_\circ = \mathcal N \left( 0, \Sigma_\circ \right)$ is given as, 
% \begin{align}
%     \Cs{r} = & \inf_{\substack{\Sigma_Z \succeq 0 \\ \BW(\Sigma_Z, \Sigma_\circ) \leq r}} \sup_{ \substack{Q \succeq 0 \\ \tr(Q) \leq B }} \thalf \log \frac{|\Sigma_Z + HQH^{T}|}{|\Sigma_Z|}
%     \label{eq:capsup2},
% \end{align}
\begin{align}
    \Cs{r} = & \inf_{\substack{\Sigma_Z \succeq 0 \\ \BW(\Sigma_Z, \Sigma_\circ) \leq r}} C_{\mathcal{N}(0, \Sigma_Z)}(B).
    \label{eq:capsup2}
\end{align}
\end{theorem}
\begin{IEEEproof}
    The proof of Theorem \ref{th:main_cap} is along the same lines as that of Theorem \ref{th:main}. We leverage the fact that Gaussian noise minimizes capacity for a given noise covariance (See Lemma \ref{lem:B1} in Appendix \ref{sec:proof_cap} below). This helps us write the compound capacity in terms of the second-order statistics of the distributions in the $\Was$ ball. The operational meaning of Gaussian being the worst-case noise is discussed in \cite{lapidoth1996nearest}. See Appendix \ref{sec:proof_cap} for details. \textcolor{black}{Also note that a Gaussian nominal distribution ensures that distributions in the $\Was$-ball have bounded $2$-nd order moments \cite[Lem. 1]{yue2022linear}.}
\end{IEEEproof}

\begin{lemma}
The compound capacity in \eqref{eq:wccap} admits a dual formulation given by
% \begin{align}
%     \Cs{r} & \= \!\inf_{\substack{P_{Z|Z_\circ} \colon  \R^{d} \mapsto \R^d \\ \mathbb{E}\norm{Z - Z_\circ}^2 \leq r^2 }} \sup_{\substack{P_{X } \colon  \R^{d} \mapsto \R^d \\ \mathbb{E}\left[\norm{X}^2\right]  \leq B}} I(X; H X + Z).
% \end{align}
\begin{align}
    \Cs{r} & \= \!\inf_{\substack{P_{Z|Z_\circ}\colon\R^{d} \mapsto \R^d \\ P_Z \in \Probc(\R^{d}) \\ \mathbb{E} \left[\norm{Z - Z_\circ}^2\right] \leq r^2 }} \sup_{\substack{P_{X } \text{ on }  \R^{d} \colon \\ \mathbb{E}\left[\norm{X}^2\right]  \leq B}} I(X; H X + Z).
    \label{eq:comp_cap_dual_lemm}
\end{align}
where the $\inf$ is achieved by a Gaussian noise distribution.
\label{lem:cdual}
\end{lemma}
\begin{IEEEproof}
The result follows from \cite[Thm. 5]{csiszar1992arbitrarily}, where the authors prove the above result for the case of AVCs, which is a more general case of the setting that we consider.
\end{IEEEproof}

Consider the compound capacity of a scalar additive channel with an unknown additive noise distribution drawn from a $\Was$-ambiguity set centered at $P_\circ = \mathcal{N}(0, \sigma^2_\circ)$. First, recall the capacity of a scalar Gaussian channel \cite[Eqn. (9.17)]{tc}:
\begin{align}
    C_{\mathcal{N}(0, \sigma^2_\circ)}( B) &= \frac{1}{2} \log\left(1 + \frac{B}{\sigma^2} \right).
    \label{eq:scalar_cap_w2}
\end{align}
We now apply Theorem \ref{th:main_cap}. Plugging \eqref{eq:bwsca} and \eqref{eq:scalar_cap_w2} into \eqref{eq:capsup2} and using the fact that logarithm is monotonic, we write the worst case capacity for $P_\circ = \normal(0,\sigma_\circ^2)$ as,
\begin{align}
C_{\mathcal{N}(0, \sigma^2_\circ)}\left( B, r \right) &= \inf_{|\sigma - \sigma_\circ| \leq r } C_{\mathcal{N}(0, \sigma^2_\circ)}( B) \\
    &= \inf_{|\sigma - \sigma_\circ| \leq r } \frac{1}{2} \log\left(1 + \frac{B}{\sigma^2} \right) \\
    &= \frac{1}{2} \log\left(1 + \frac{B}{(\sigma + r)^2} \right).
\end{align}

Naturally, increasing the radius of ambiguity $r$ diminishes the compound channel capacity. This effect can be observed in Fig \ref{fig:drccf}.

\section{Proof of Theorem \ref{th:main}}
\label{sec:mainproof}
\subsection{Useful Results}

If the source distribution $X \sim P_\circ$ is known, \eqref{eq:wcrdf} reduces to the standard RDF \cite[Eqn. (10.12)]{tc}
\begin{align}
   R_{P_\circ}(D) = \RDg{0} = \inf_{\substack{\EP \colon  \R^{d} \mapsto \R^d \\ \mathbb{E}\left[\|X - \widehat{X}\|^2\right]  \leq D}} I(X; \widehat{X}).
   \label{eq:RDFP0}
\end{align}

The next lemma brings to light the significance of the Gaussian RDF in \eqref{eq:gaussian_rdf} as the worst-case RDF among sources with the same covariance matrix.

\begin{lemma}[Gaussian is the hardest to encode, {\cite[Eqn. (13)]{kolmogorov1956shannon}}]
    Assume that the source distribution $P_\circ$ has a covariance $\Sigma_\circ \in \Sym_{+}^d$. The RDF $R_{P_\circ}(D)$ satisfies
    \begin{equation}
    R_{P_\circ}(D) \leq R_{\normal(0,\Sigma_\circ)}(D),
    \end{equation} 
    % where $\KL(P||Q)$ is the KL divergence between distributions $P$ and $Q$
    and the upper bound is attained by $P_\circ = \mathcal{N}\left(0, \Sigma_\circ\right)$.\label{th:worst_case_rdf_vector}
\end{lemma}
\begin{IEEEproof}
  This result is due to Kolmogorov \cite[Eqn. (13)]{kolmogorov1956shannon}. We provide a short proof in Appendix \ref{proof:worst_case_rdf_vector} for completeness.  
\end{IEEEproof} 
Noting that 
\begin{align}
    \Was(P,P_\circ)^2 \leq r  \Longleftrightarrow \E[\|X - X_\circ\|^2] \leq r,
\end{align}
we rewrite the dual formulation of the compound RDF in Lemma~\ref{lem:rdfdual} as,
\begin{equation}
\RDg{r}  \= \!\sup_{\substack{P \in\Prob(\R^d) \colon \\ \Was(P,\, P_\circ) \leq r}} R_{P}(D).
\label{eq:rdfduala}
\end{equation}

\subsection{Proof of Theorem~\ref{th:main}: upper bound}
\label{sec:rdfupper}

Consider two distributions $P_\circ, P\in\Prob(\R^d)$ with means $\mu_\circ,\mu\in\R^d$ and covariances $\Sigma_\circ,\Sigma \in \Sym_{+}^{d}$, respectively. The $\Was$-distance between them satisfies the Gelbrich bound \cite[Thm. 2.1]{gelbrich},
\begin{align}
    \Was(P,P_\circ)^2 &\geq \BW(\Sigma, \Sigma_\circ)^2 \+ \norm{\mu_\circ\-\mu}^2, \label{eq:gelbrich}
\end{align}
where equality is attained if both $P_\circ$ and $P$ are Gaussian distributions. It follows that the $\Was$-distance between distributions upper-bounds the Bures-Wasserstein distance between their covariance matrices:
\begin{align}
    \Was(P,\, P_\circ) & \geq \BW(\Sigma,\, \Sigma_\circ).
    \label{eq:w2bwineq}
\end{align}
Applying \eqref{eq:w2bwineq} to \eqref{eq:rdfduala}, we obtain an upper bound on the worst-case RDF as:
\begin{align}\label{eq:rdf_ineqa_new}
     \RDg{r} &\leq  \sup_{\substack{P \in\Prob(\R^d) \colon \\ \BW(\Sigma,\, \Sigma_\circ) \leq r}}\ R_{P}(D) \\
     &\leq \sup_{\BW(\Sigma, \Sigma_\circ) \leq r} R_{\normal(0,\Sigma)}(D) \label{eq:rdf_ineqa},
\end{align}
where \eqref{eq:rdf_ineqa} is by Lemma~\ref{th:worst_case_rdf_vector}.
Plugging \eqref{eq:gaussian_rdf} into the right side of \eqref{eq:rdf_ineqa} yields the  $\leq$ direction of \eqref{eq:main}.

\subsection{{Proof of Theorem~\ref{th:main}: lower bound}}
\label{sec:rdflower}
By Lemma~\ref{th:worst_case_rdf_vector}, equality is achieved in \eqref{eq:rdf_ineqa} by $P = \normal(0,\Sigma)$. Since both $P_\circ$ and $P$ are Gaussian, equality is achieved in \eqref{eq:w2bwineq}, and, by extension, in \eqref{eq:rdf_ineqa_new} as well.

% \section{Coding Theorems}
% \input{sections/coding_theorems}

%\section{Examples and Simulations}
%\input{sections/examples}

% \section{Worst Case Channel Capacity}
% \input{sections/capacity}

\section{Conclusion}
In this paper, we studied Sakrison's \cite{sakrison1969rate} compound RDF and Blackwell's compound capacity \cite{blackwell1959capacity}, focusing on a scenario where the source / noise distribution belongs to a Wasserstein-2 ambiguity set. Our key findings include the identification of the Gaussian distribution as the worst-case scenario for encoding within this set (Theorem~\ref{th:main}, Theorem~\ref{th:main_cap}). The compound RDF (Theorem~\ref{th:main}) and capacity (Theorem~\ref{th:main_cap}) are expressed in terms of the covariance matrices of Gaussian distributions.
Future work could explore ambiguity sets defined using distances beyond Wasserstein-2, tradeoffs between coding and learning the distribution to decrease the size $r$ of the ambiguity set, extensions to multiterminal settings, and to causal source and channel coding.

%%%%%%
%% Appendix:
%% If needed a single appendix is created by
%%
%\appendix
%%
%% If several appendices are needed, then the command
%%
% \appendices
%%
%% in combination with further \section commands can be used.
%%%%%%

% \section*{Acknowledgment}

% We are indebted to Michael Shell for maintaining and improving
% \texttt{IEEEtran.cls}. 

%%%%%%
%% To balance the columns at the last page of the paper use this
%% command:
%%
%\enlargethispage{-1.2cm} 
%%
%% If the balancing should occur in the middle of the references, use
%% the following trigger:
%%
% \IEEEtriggeratref{4}
%%
%% which triggers a \newpage (i.e., new column) just before the given
%% reference number. Note that you need to adapt this if you modify
%% the paper.  The "triggered" command can be changed if desired:
%%
%\IEEEtriggercmd{\enlargethispage{-20cm}}
%%
%%%%%%

%%%%%%
%% References:
%% We recommend the usage of BibTeX:
%%
% \vk{Please carefully go through all the references, make sure they are in the same format, fix capitalization issues, and provide publication month.} \vm{Fixed the important capital words.}

\bibliographystyle{IEEEtran}
\bibliography{references}

\newpage
\clearpage
\appendix
% \begin{center}
% {\huge Appendix}
% \end{center}
\appendices
\section{Proof of Lemma \ref{th:worst_case_rdf_vector}}
\label{proof:worst_case_rdf_vector}

Without loss of generality, assume that $X$ is zero-mean. We first state the following useful lemma, which extends \cite[Thm. 5.11]{polyanskiy2022information} to a vector channel.
\begin{lemma}[Gaussian input maximizes the mutual information in an AWGN channel]
    Let $X$ be a random channel input in $\R^d$ with known covariance %$\Sigma_X \in \R^{d\times d}$, 
    {$\Sigma \in \Sym^{d}_{+}$}
    and let $A\in \R^{d\times d}$ be a fixed channel matrix. Let $Z_g \sampled \normal(0, \Sigma_Z)$ be the additive Gaussian channel noise independent of $X$, with known covariance %$\Sigma_Z \in \R^{p\times p}$
    {$\Sigma_Z \in \Sym^{d}_{+}$}. We have that;
    \begin{equation}
        I(X,\; AX + Z_g) \leq I(X_g, \; AX_g + Z_g),
        \label{eq:app_A_gaussian}
    \end{equation}
    where $X_g \sampled \normal(0, \Sigma)$ is independent of $Z_g$, and equality holds if $X \sim \normal(0, \Sigma)$.
    \label{lem:gwc}
\end{lemma}
\begin{IEEEproof}
The mutual information is given by,
    \begin{align}
         I(X,\, AX + Z_g) &= h(AX+Z_g) - h(AX+Z_g | X), \\
         &= h(AX+Z_g) - h(Z_g).
    \end{align}
The second term in the last equality does not depend on $X$, while the first term, by the maximum entropy property of Gaussian random variables \cite[Thm. 8.6.5]{tc}, is bounded above by
\begin{equation}
     h(AX+Z_g) \leq
    \frac{1}{2}\log \left|2\pi \e \,\widehat{\Sigma} \right|,
    \label{eq:gaussian_maximises_entropy}
\end{equation}
where $\widehat{\Sigma} \defeq \Cov(AX+Z_g) = A\Sigma A^T + \Sigma_Z$ is the covariance of matrix of $AX+Z_g$. This upper bound is achieved by a Gaussian input $X_g \sampled \mathcal{N}(0, \Sigma)$ independent of the channel noise $Z_g$.
\end{IEEEproof}

Consider the RDF for a known source distribution $P_X$ given in \eqref{eq:RDFP0}. Restricting the infimization to linear mappings of the form
\begin{align}
    \widehat{X} = AX + Z_g,
\end{align}
where $A \in \mathbb{R}^{d \times d}$, and $Z_g \sim \mathcal{N}(0, \Sigma_Z)$ is independent of $X$, we get an upper bound on $R_{P_X}(D)$ as,
\begin{align}
    % \RDp = & 
    R_{P_X}(D) &\leq \inf_{\substack{A \in \mathbb{R}^{d \times d}, \Sigma_Z \succeq 0 \colon \\ \mathbb{E}\left[ \norm{X -\widehat{X}}^2 \right]  \leq D}} I(X; AX + Z_g) \label{eq:rdf_ineq_app_a} \\
    & \leq \inf_{\substack{A \in \mathbb{R}^{d \times d}, \Sigma_Z \succeq 0 \colon \\ \mathbb{E}\left[ \norm{X_g -\widehat{X}_g}^2 \right]  \leq D}} I(X_g; AX_g + Z_g) \label{eq:rdf_ineq_app_b} \\
    &= R_{\mathcal{N}(0, \Sigma)}(D), \label{eq:rdf_ineq_app_b_last}
\end{align}
where $X_g \sim \mathcal{N}(0, \Sigma)$ and $\widehat{X}_g \coloneqq AX_g + Z_g$. Inequality \eqref{eq:rdf_ineq_app_b} is by Lemma \ref{lem:gwc}.
Since the condition $\mathbb{E}\left[ \norm{X_g -\widehat{X}_g}^2 \right]  \leq D$ can be written as $\tr( (A \- I) \Sigma(A \- I)^T \+ \Sigma_Z) \leq D$, 
and the mutual information between two Gaussian random vectors can be written as,
\begin{align}
    I(X_g; AX_g + Z_g) = \frac{1}{2} \log  \frac{\left|A \Sigma A^T+\Sigma_Z\right|}{\left|\Sigma_Z\right|},
    \label{eq:gaussian_mutual}
\end{align}
the right side of \eqref{eq:rdf_ineq_app_b} is equal to the right side of \eqref{eq:gaussian_rdf}. It remains to show \eqref{eq:rdf_ineq_app_b_last}. Applying the argument in \eqref{eq:rdf_ineq_app_a}, \eqref{eq:rdf_ineq_app_b} to $P_X = \mathcal{N}(0, \Sigma)$ yields the $\geq$ in \eqref{eq:rdf_ineq_app_b_last}. To show $\leq$, fix any $P_{\widehat{X}|X_g}$. Using standard arguments, \cite[Proof of Th. 10.3.2]{tc}, we have,
    \begin{align}
        I(X_g, \widehat{X}) &= h(X_g) - h(X_g | \widehat{X}) \\
        &= \frac{1}{2} \log \left|2 \pi \e  \Sigma \right| - h(X_g - \widehat{X} | X_g) \\
        &\geq \frac{1}{2} \log \left|2 \pi \e  \Sigma \right| - h(X_g - \widehat{X}) \label{eq:gworst_0} \\
        &\geq \frac{1}{2} \log \left|2 \pi \e  \Sigma \right| - \frac{1}{2} \log \left|2 \pi \e  \Sigma_{X_g - \widehat{X}} \right| \label{eq:gworst_1},
    \end{align}
where $\Sigma_{X_g - \widehat{X}}$ is the covariance of $X_g - \widehat{X}$. Here, \eqref{eq:gworst_0} holds because conditioning decreases entropy \cite[Thm. 8.6.1 Cor. 2]{tc}, and \eqref{eq:gworst_1} is due to \eqref{eq:gaussian_maximises_entropy}. The expression on the right-hand side of \eqref{eq:gworst_1} is the mutual information between the jointly Gaussian pair $(X_g, \widehat{X})$. It follows that $R_{\mathcal{N}(0, \Sigma)}(D)$ is lower bounded by the right side of \eqref{eq:rdf_ineq_app_b}.

\section{Proof of Theorem~\ref{th:main_cap}}
\label{sec:proof_cap}

We first state the following useful lemmas. In Lemma \ref{lem:multi_cap}, we state the capacity of a Gaussian vector channel. To get Lemma \ref{lem:multi_cap}, we slightly modify \cite[Thm. 9.1]{el2011network} for our use case. In Lemma \ref{lem:B1}, we show that for an additive channel, Gaussian noise, among all noises with the same covariance, minimizes the channel capacity. A similar result in the setting of stochastic processes is established in \cite{ihara1978capacity}.
\begin{lemma}[Capacity of an AWGN channel]
\label{lem:multi_cap}
Consider the channel coding setting defined in the first paragraph of Section \ref{sec:channel_coding}. The capacity-cost function for a $d$-dimensional multivariate Gaussian noise vector $Z_g \sim \mathcal{N}(0, \Sigma_Z)$ is given by, 
\begin{align}
    C_{\mathcal{N}(0, \Sigma_Z)}(B) = & \sup_{ \substack{\Sigma \succeq 0 \colon \\ \text{tr}(\Sigma) \leq B }} \log \frac{|\Sigma_Z + H\Sigma H^{T}|}{|\Sigma_Z|}.
    \label{eq:Cgauss}
\end{align}
\end{lemma}
\begin{IEEEproof}
The capacity for $Z_g \sim \mathcal{N}(0, I)$, is given by \cite[Thm. 9.1]{el2011network},
\begin{align}
    C_{\mathcal{N}(0, I)}(B) = & \sup_{ \substack{\Sigma \succeq 0 \colon \\ \text{tr}(\Sigma) \leq B }} \log |I + H\Sigma H^{T}|.
\end{align}
The capacity for $Z_g \sim \mathcal{N}(0, \Sigma_Z)$ can be obtained by considering the channel matrix $\Tilde{H} = \Sigma_Z^{-1/2}H$ \cite[Rem. 9.1]{el2011network}.
\begin{align}
    C_{\mathcal{N}(0, \Sigma_Z)}(B) = & \sup_{ \substack{\Sigma \succeq 0 \colon \\ \text{tr}(\Sigma) \leq B }} \log |I + \Sigma_Z^{-1/2}H\Sigma H^{T}\Sigma_Z^{-T/2}|.
    \label{eq:idk}
\end{align}
Multiplying $I + \Sigma_Z^{-1/2}HQH^{T}\Sigma_Z^{-T/2}$ by $\Sigma_Z^{1/2}$ on the left and $\Sigma_Z^{T/2}$ on the right, and dividing the argument of $\log$ in \eqref{eq:idk} by $|\Sigma_Z|$, we get \eqref{eq:Cgauss}.
\end{IEEEproof}

\begin{lemma}[Gaussian noise minimizes the channel capacity of an additive channel]
\label{lem:B1}
    Let $Z$ be a random channel noise in $\R^d$ with known covariance $\Sigma_Z \in \R^{d\times d}$
    and let $H\in \R^{d\times d}$ be a fixed channel matrix. Let $Z_g \sampled \normal(0, \Sigma_Z)$ be the additive Gaussian channel noise, with the same covariance as $Z$, independent of the channel input $X$. Then,
    \begin{align}
        C_{\mathcal{N}(0, \Sigma_Z)}(B) \leq C_{P}(B),
        \label{eq:41}
    \end{align}
and equality is achieved by $P = \mathcal{N}(0, \Sigma_Z)$.
\end{lemma}
\begin{IEEEproof}
    Consider $X_g \sim \mathcal{N}(0, \Sigma)$ where $\Sigma$ is the covariance of $X$. Using \eqref{eq:app_A_gaussian} and then \cite[Lem. II.2]{diggavi2001worst}, we have the following set of inequalities,
    \begin{align}
        I(X,\; HX + Z_g) &\leq I(X_g, \; HX_g + Z_g), \label{eq:ineqsss0}\\
        &\leq I(X_g,\; HX_g + Z).
        \label{eq:ineqsss}
    \end{align}
    Now note that,
    \begin{align}
        \sup_{\substack{ \Sigma \succeq 0\colon\\\mathbb{E}\left[\norm{X_g}^2\right]  \leq B}} I(X_g,\; HX_g + Z) \leq \sup_{\substack{P \text{ on } \R^d\colon \\  \mathbb{E}\left[\norm{X}^2\right]  \leq B}} I(X, HX + Z).\label{eq:ineqsss90}
    \end{align}
     The expression on the right-hand side of \eqref{eq:ineqsss90} is $C_P(B)$. The expression on the left-hand side of \eqref{eq:ineqsss90} is lower bounded by $C_{\mathcal{N}(0, \Sigma_Z)}(B)$. Indeed, After applying inequalities \eqref{eq:ineqsss} and then \eqref{eq:ineqsss0} to lower-bound the left side of \eqref{eq:ineqsss90} and taking supremum over all input distributions leads to \eqref{eq:Cgauss}.
\end{IEEEproof}

\textit{Proof of Theorem~\ref{th:main_cap}:}
\subsubsection{Converse}
Consider the dual formulation of the compound capacity \eqref{eq:comp_cap_dual_lemm} and a 
noise distribution in the $\Was$ ball $P$ with mean $\mu$ and covariance $\Sigma$. We re-write the dual formulation as,
\begin{align}
\Cs{r}  \= \!\inf_{\substack{P \in\Probc(\R^d) \colon \\ \Was(P,\, P_\circ) \leq r}}C_P(B).
\label{eq:app_cap_dual}
\end{align}
By the Gelbrich bound \cite[Thm. 2.1]{gelbrich} we have \eqref{eq:w2bwineq}.
Applying  \eqref{eq:w2bwineq} to \eqref{eq:app_cap_dual}, we obtain a lower bound on the compound capacity as$\colon$
\begin{align}
    \Cs{r} &\geq \!\inf_{\substack{P \in\Probc(\R^d) \colon \\ \BW(\Sigma, \Sigma_\circ) \leq r}}C_P(B) \label{eq:gaussian_minimises_cap0} \\
    &\geq  \!\inf_{\substack{P \in\Probc(\R^d) \colon \\ \BW(\Sigma, \Sigma_\circ) \leq r}}C_{\mathcal{N}(0, \Sigma)}(B), \label{eq:gaussian_minimises_cap}
\end{align}
where \eqref{eq:gaussian_minimises_cap} is by Lemma \ref{lem:B1}. Plugging \eqref{eq:Cgauss} into the right side of \eqref{eq:gaussian_minimises_cap} yields the $\geq$ direction of \eqref{eq:capsup2}.
\subsubsection{Achievability}
Equality in \eqref{eq:gaussian_minimises_cap} is achieved by Gaussian $P$ \eqref{eq:41}. Since we assume that $P_{\circ}$ is Gaussian, equality is achieved in \eqref{eq:w2bwineq}, and, by extension, in \eqref{eq:gaussian_minimises_cap0} as well.

%%
%% where we here have assumed the existence of the files
%% definitions.bib and bibliofile.bib.
%% BibTeX documentation can be obtained at:
%% http://www.ctan.org/tex-archive/biblio/bibtex/contrib/doc/
%%%%%%

%% Or you use manual references (pay attention to consistency and the
%% formatting style!):
% \begin{thebibliography}{9}

% \bibitem{Laport:LaTeX}
% L.~Lamport,
%   \emph{\LaTeX: A Document Preparation System,} 
%   Addison-Wesley, Reading, Massachusetts, USA, 2nd~ed., 1994. 

% \bibitem{GMS:LaTeXComp}
% F.~Mittelbach, M,~Goossens, J.~Braams, D.~Carlisle, and
% C.~Rowley, \emph{The {\LaTeX} Companion,} Addison-Wesley,
% Reading, Massachusetts, USA, 2nd~ed., 2004.

% \bibitem{oetiker_latex}
% T.~Oetiker, H.~Partl, I.~Hyna, and E.~Schlegl, \emph{The Not So Short
%   Introduction to {\LaTeX2e}}, version 5.06, Jun.~20, 2016. [Online].
%   Available: \url{https://tobi.oetiker.ch/lshort/}

% \bibitem{typesetmoser}
% S.~M. Moser, \emph{How to Typeset Equations in {\LaTeX}}, version 4.6,
%   Sep. 29, 2017. [Online]. Available:
%   \url{http://moser-isi.ethz.ch/manuals.html#eqlatex}

% \bibitem{IEEE:pdfsettings}
% IEEE, \emph{Preparing Conference Content for the IEEE Xplore Digital
%   Library.} [Online]. Available:
%   \url{http://www.ieee.org/conferences_events/conferences/organizers/pubs/preparing_content.html}

% \bibitem{IEEE:AuthorToolbox}
% IEEE, \emph{Author Digital Toolbox.} [Online.] Available:
%   \url{http://www.ieee.org/publications_standards/publications/authors/authors_journals.html}

% \end{thebibliography}

\end{document}